\documentclass[twocolumn,preprintnumbers,superscriptaddress,endnote,nofootinbib,aps,prd,floatfix]{revtex4}

\usepackage{color}

\usepackage{footmisc,multirow}
\usepackage{subfigure}
\usepackage{amsmath,slashed}
\usepackage{graphicx}
\usepackage{hyperref}
\usepackage[hyphenbreaks]{breakurl}

\newcommand{\rig}{\rightarrow}

\newcommand{\be}{\begin{eqnarray*}}
\newcommand{\ee}{\end{eqnarray*}}
\newcommand{\gl}[1]{(\ref{#1})}

\newcommand{\bee}{\begin{eqnarray}}
\newcommand{\eee}{\end{eqnarray}}
\newcommand{\beeq}{\begin{equation}}
\newcommand{\eeeq}{\end{equation}}
\newcommand{\gev}{{\rm{GeV}}}

\begin{document}

\author{Christoph Englert} 
\email{christoph.englert@durham.ac.uk}
\affiliation{Institute for Particle Physics Phenomenology, Department
  of Physics,\\Durham University, DH1 3LE, United Kingdom}
\author{Dorival Gon\c{c}alves Netto}
\email{d.goncalves@thphys.uni-heidelberg.de}
\affiliation{Institut f\"ur Theoretische Physik, Universit\"at Heidelberg, 69120
  Heidelberg, Germany}
\author{Michael Spannowsky}
\email{michael.spannowsky@durham.ac.uk}
\affiliation{Institute for Particle Physics Phenomenology, Department
  of Physics,\\Durham University, DH1 3LE, United Kingdom}
\author{John Terning}
\email{terning@physics.ucdavis.edu}
\affiliation{Department of Physics, University of California, Davis, CA 95616, USA}

\title{Constraining the Unhiggs with LHC data}

\begin{abstract}
  Recent measurements by the ATLAS and CMS experiments have excluded
  the Standard Model Higgs boson in the high mass region, even if it
  is produced with a significantly smaller cross section than expected. The bounds
  are dominated by the non-observation of a signal in the clean
  gold-plated mode $h\to ZZ\to 4\ell$ and, hence, are directly
  related to the special role of the Higgs in electroweak symmetry
  breaking. A smaller cross section in comparison to the Standard
  Model is expected if the Higgs is realized as an unparticle in the
  Unhiggs scenario. With the LHC probing
  $\sigma/\sigma^{\text{SM}}<1$, we can therefore reinterpret the
  $h\to ZZ\to 4\ell$ exclusion limits as bounds on the Unhiggs'
  scaling dimension. Throughout the high Higgs mass range, where we
  expect a large signal in the presence of the Standard Model Higgs
  for the 2011 ATLAS and CMS data sets,  the observed
  limits translate into mild bounds on the Unhiggs scaling dimension
  in the high mass region.
\end{abstract}

\maketitle

\section{Introduction}
\label{sec:intro}
Recent results on Standard Model Higgs production at the LHC
\cite{ATLASCouncil, CMSCouncil,atlaszz,cmszz,recent} have gained lots
of attention throughout the high energy physics community. On the one
hand, this is due to the tantalizing hints for a light Higgs boson
around $m_h\simeq 125~\gev$. The implications of the observed excess have already been
discussed in the literature~\cite{data}, yet, it is too early to claim
that the Higgs has been found and we might well observe a statistical
fluctuation~\cite{fitdata}.

On the other hand, the 95\% confidence level bounds published by ATLAS
and CMS state that if a Higgs-like resonance is to be realized at
higher masses, the Higgs boson is significantly underproduced or has
suppressed branching fractions to the (partially) visible decay
channels. This has triggered some effort in how to reconcile one or
more heavy Higgs bosons in the light of this heavily constraining data
through modified production cross sections~\cite{prod},
decays~\cite{decay}, or combinations of both~\cite{proddec}.

The dominant channel which drives the exclusion bounds in the high
Higgs mass regime is the clean Higgs decay channel to four leptons via
two $Z$ bosons~\cite{atlaszz,cmszz,nlomat}. This so-called ``gold-plated''
mode allows a great deal of Higgs ``spectroscopy'' as soon as the
decay channel $h\to ZZ$ opens up. Statics in this channel is limited
due to the branching ratios of the $Z$ bosons to the light and clean
leptons $e^\pm,\mu^\pm$, but it benefits from a large branching ratio
$h\to ZZ$ for heavy Higgs masses $\Gamma(h\to ZZ)\sim m_h^3/m_Z^2$.
The dominant partial decay width to longitudinal $W,Z$ is a direct
consequence of electroweak symmetry breaking.  The purely leptonic
final state of $h\to 4\ell$ can be fully reconstructed and, hence, its
merits range from line-shape measurements~\cite{lineshape} to
obtaining spin and ${\cal{CP}}$ of the reconstructed
excess~\cite{spincp}. Of similar importance in the high Higgs mass
region at the 14~TeV LHC are the semi-hadronic $ZZ$ decay modes for
boosted kinematics~\cite{hadzz}.

Avoiding a large branching ratio $h\to ZZ$, or in general $h\to VV$,
$V=W^\pm,Z$, for heavy Higgs particles is theoretically challenging
unless we allow for non-perturbative strong
couplings~\cite{Englert:2012dq}. The reason is that ordinary
perturbative ${\cal{O}}(1)$ interactions of a minimally extended Higgs
sector, {\emph{i.e.}} the ones which do not arise from spontaneous
symmetry breaking, cannot compete against the fast-growing partial
decay width of the Higgs $\sim m_h^3$. Higgs-portal type interactions
are pushed into a non-perturbative regime by requiring a vanishing
$h\to ZZ $ phenomenology~\cite{decay}.

One way to reconcile a vanishing Higgs phenomenology in $h\to ZZ \to
4\ell$ in a controlled way is by turning to strongly coupled theories
in the `t Hooft limit~\cite{tHooft:1973jz}. A playground, which serves
as a model-building dictionary, is the AdS/CFT
correspondence~\cite{adscft,Csaki:2008dt} of Randall-Sundrum
models~\cite{rs}. While RS I type models predict a tower of scalar
and/or vectorial resonances which can be tackled with standard search
strategies in the $VV$ final
states~\cite{Agashe:2006hk,Contino:2011np}, the phenomenology of RS II
models can be fundamentally different. Once put into a realistic form
by introducing a weak breaking of conformal invariance via a
modulation of the AdS metric towards the infrared~\cite{adsgeorgi},
such models can be interpreted as ``Unparticles'' in Georgi's
language~\cite{georgi}. It has been shown that such a non-local object
can, in fact, be responsible for electroweak symmetry breaking,
restoring the good high energy behavior of longitudinal gauge boson
scattering~\cite{unhiggs}, while it behaves very much like an ordinary
Higgs in electroweak precision tests \cite{david10}. In essence, the
model's gauge sector at low energies is similar to the SM, while this
is not necessarily true for the Unhiggs-fermion
sector~\cite{Englert:2012dq}. Therefore every channel which allows
cross-talk between the fermion and the gauge boson sector is a strong
probe of such a mechanism of electroweak symmetry breaking.

The production of the Unhiggs from gluon fusion and its subsequent
decay precisely serves this purpose. Given the recently observed
underproduction of Higgs-like states in the $ZZ$ channel it is
possible to formulate bounds on Unhiggs-symmetry breaking, which is
the purpose of this paper. As it turns out, the gold-plated mode is
perfectly suited for such a reanalysis, since systematics for heavy
Higgses are sufficiently small in the fully reconstructed final
state. Note that this is vastly different from $h\to ZZ\to
\ell^+\ell^-{\slashed{E}}_T$, which drives the Higgs exclusion for
very heavy Higgs masses~\cite{recent}. Systematic uncertainties in
these channels limit a straightforward re-application of the existing
strategies pursued by ATLAS and CMS.

\bigskip

We organize this work as follows: Sec.~\ref{sec:model} provides a
recap of the model's properties to make this work self-contained. In
Sec.~\ref{sec:analysis} we first validate our analysis strategy
against the results of Refs.~\cite{atlaszz,cmszz}. In particular we
show that we can reproduce the experiments' exclusion bounds in the
Higgs mass region we are interested in. We subsequently compute the
exclusion bounds on the Unhiggs model's parameters that are implied by
the data of Refs.~\cite{atlaszz,cmszz} in Sec.~\ref{sec:bounds}. We
give our conclusions in Sec.~\ref{sec:sum}.

\section{The Model}
\label{sec:model}
The gauge interactions of the Unhiggs field $H$ follow from the
effective lagrangian~\cite{unhiggs}
\begin{equation}
\label{eq:switchoff}
  {\cal{L}} \supset H^\dagger \left( D^\mu D_\mu + \mu^2 \right)^{2-d}H \,,
\end{equation}
where $D_\mu$ denotes the familiar $SU(2)_L\times U(1)_Y$ gauge
covariant derivative, $d<2$ is the Unhiggs field's scaling dimension
and $\mu$ is the infrared cut-off of the conformal
sector. Consequently, the gauge interactions of the Unhiggs can be
dialed away by increasing $d>1$. The naive growth with the center of
mass energy of the longitudinal gauge boson scattering amplitude,
which would eventually lead to unitarity violation if left un-tamed,
is cured by non-local interactions~\cite{unhiggs}. Similar gauge
cancellations in massive quark annihilation to longitudinal gauge
bosons constrain the Unhiggs scaling dimension $d\lesssim
1.5$~\cite{Englert:2012dq}. $d \gtrsim 1.5$ implies modifications of
the massive fermion sector by inducing non-local contributions. These
can eventually alter also the Higgs phenomenology for low masses in
{\emph{e.g.}} loop-induced $h\to \gamma \gamma$, where the $125~\gev$
excess is observed. For small $d\simeq 1$ the effective theory that
follows from Eq.~\gl{eq:switchoff} and the SM Yukawa sector is
unitarity-conserving, and results in a similar Higgs phenomenology for
Higgs masses below the $Z$ and $W^\pm$ thresholds.

Since the resulting Higgs field is manifestly non-local as a
consequence of Eq. \gl{eq:switchoff}, so are the longitudinal gauge
bosons, {\emph{i.e.}} the would-be Nambu Goldstone bosons in unitary
gauge. The Higgs two-point function is given by a pole at what would
be interpreted as the physical Higgs particle (dubbed ``Unhiggs'' in
the following) and a branch cut above the conformal symmetry breaking
scale $\mu$. The impact of this continuum is, however,
phenomenologically irrelevant at the LHC for small $d>1$ since it is
difficult to access the gauge boson's polarizations in a clean way in the
context of Higgs production~\cite{Englert:2012dq,Han:2009em}. The
resulting contribution to the cross section of the order of a few
percent in $pp\to h\to ZZ+X$, is difficult to isolate from
background uncertainties, especially at the given luminosity
${\cal{L}}\simeq 5~{\rm{fb}}^{-1}$, which predicts only a couple of
events in the decay leptons invariant mass tails. Hence search
strategies will be sensitive to the modifications of the Unhiggs with
respect to the SM Higgs.

\begin{figure*}[t]
  \begin{center}
    \parbox{0.4\textwidth}{ \subfigure[][\label{fig:exirefa}]{
      \includegraphics[width=0.4\textwidth]{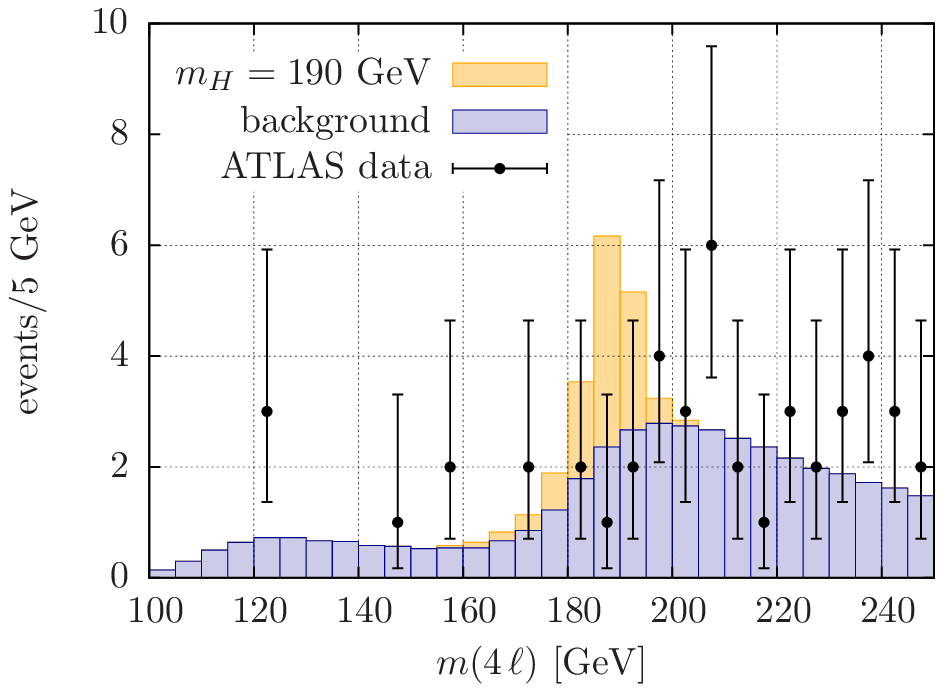}
    }}\hspace{1cm}
  \parbox{0.4\textwidth}{ \subfigure[][\label{fig:exirefb}]{
      \includegraphics[width=0.4\textwidth]{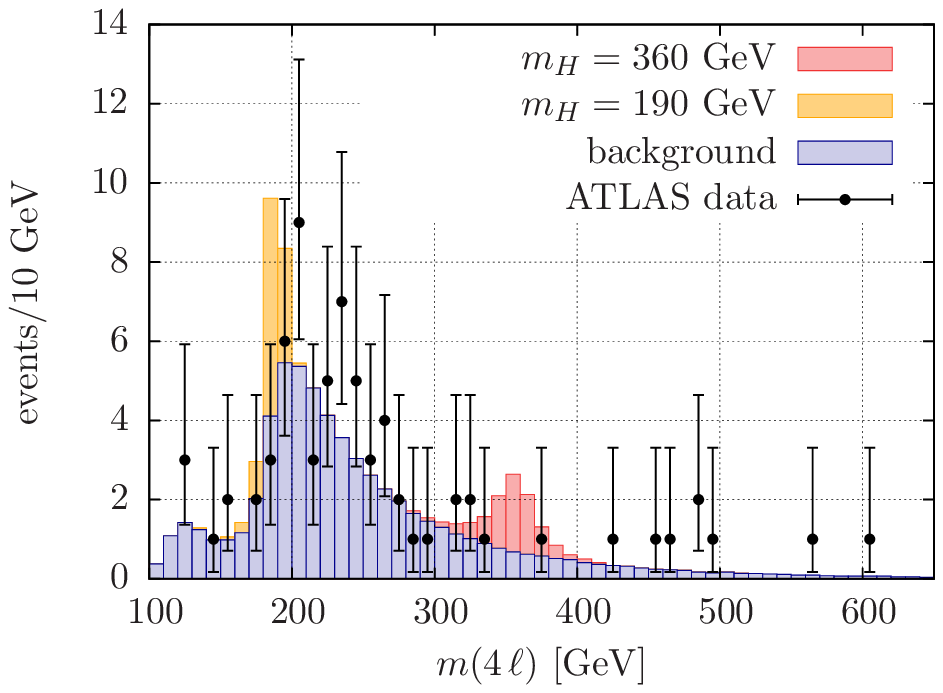}
    }}\\[0.3cm]
  \parbox{0.44\textwidth}{
    \subfigure[][\label{fig:exirefc}]{
      \includegraphics[width=0.4\textwidth]{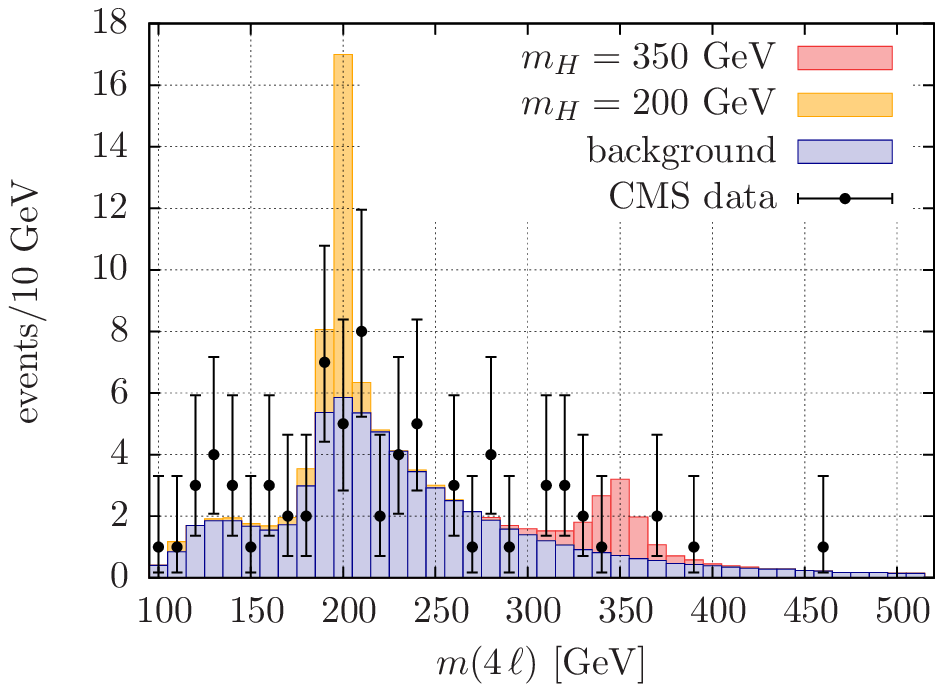}
    }}\hspace{1cm}
  \parbox{0.4\textwidth}{ \vspace{-2cm}\caption{\label{fig:exiref}
      ATLAS distributions and data (a),~(b) (taken from
      Ref.~\cite{atlaszz}) and CMS distributions and data (c) (taken
      from Ref.~\cite{cmszz}), which enter the hypothesis tests of
      Tab.~\ref{tab:comp}.} }
\end{center}
\end{figure*}

Due to the modifications of the Higgs sector that arise from
Eq.~\gl{eq:switchoff} after electroweak symmetry breaking, both the
partial decay width ({\emph{i.e.}} the line shape) and the production
cross section are modified. More precisely, the Unhiggs propagator
reads \cite{unhiggs}
\begin{equation}
  \label{eq:prop}
  \Delta_H=-\frac{i}{ (\mu^2 -q^2)^{2-d} - (\mu^{2} -m_h^{2})^{2-d} }\,.
\end{equation}
where $m_h$ is the pole mass. The top-Yukawa coupling for $d\lesssim
1.5$ follows from ${\cal{L}}\supset ({\lambda_t/\sqrt{2}})\,
({v^{d}/\Lambda^{d-1}})\,\bar t_R t_L + {\rm{h.c.}}$, where $\Lambda$
is the model's cut-off scale (see~Refs.~\cite{Englert:2012dq,unhiggs}
for further details).
\begin{subequations}
  \begin{widetext}
    \begin{align}
      \label{eq4a}
      \frac{\Gamma^{\rm{Unh}}}{\Gamma^{\rm{SM}}} 
        \simeq {(\mu^2)^{d-1}\over 2-d}\left( 
          \frac{(\mu^2)^{2-d}-(\mu^2-m_h^2)^{2-d}}{m_h^2} \right)^2  
        {-\pi {\cal{A}}_d \over 2\pi \sin( \pi d ) } {(  \mu^2-m_h^2)^{d-1}\over 2-d}\,,\\
        \label{eq4b}
        {\sigma^{\rm{Unh}} (gg\rig H \rig VV)\over 
          \sigma^{\rm{SM}}(gg\rig H \rig VV)}  
        \simeq \left| {\left(1 -
              {m_h^2\over q^2}\right) }\frac {[(\mu^2 - q^2)^{2 - d} - 
            \mu^{4 - 2d}]}{(\mu^2 - q^2)^{2 - d}-(\mu^2 -
            m_h^2)^{2 - d}}
        \right|^2_{q^2=\sqrt{s}}\,,
      \end{align}
  \end{widetext}
 where
  \begin{equation}    
    \label{eq4c}
    {\cal{A}}_d={16\pi^{5/2}\over (2\pi)^{2d}} {\Gamma(d+1/2)\over \Gamma(d-1) \Gamma(2d)}\,.
  \end{equation}
\end{subequations} 
The above modification of width stems from the pole contribution and
its decay to SM particles. This should be contrasted to an additional
imaginary part that the propagator Eq.~\gl{eq:prop} can pick
up. Eq.~\gl{eq4b} gives an approximation for the pole contribution in
terms of cross section times branching ratio. In the final analysis we
use the full off-shell propagator including imaginary parts in
Sec.~\ref{sec:bounds}, {\emph{i.e.}} the one that arises from the
conformal structure of the propagator and the one that arises from the
decay of the $m_h$ state. Doing so, we recover the SM propagators,
widths and cross sections upon taking the limit $d\to 1$. This is also
clear from Eq.~\gl{eq4a}-\gl{eq4c}, for $d\to 1$ we have
$\sigma^{\rm{Unh}}/\sigma^{\rm{SM}},\Gamma^{\rm{Unh}}/{\Gamma^{\rm{SM}}}\to
1$.

The $pp\to ZZ+X\to 4\ell+X$ channel is potentially sensitive to both
of these modifications in the heavy mass region $m_h\gtrsim 200~\gev$,
especially because the observed Higgs width is dominated by the
physical one~\cite{atlaszz,cmszz}. For light Higgs particles the width
is dominated by the detector resolution, also in the gold-plated $h\to ZZ$
decay mode (see {\emph{e.g.}}~\cite{searches}).

\vspace{-0.2cm}
\section{Elements of the Analysis}
\label{sec:analysis}

\subsubsection*{Statistics}
To formulate exclusion bounds on the Unhiggs model, we apply a binned
log-likelihood ratio hypothesis test as formulated during the LEP2 era
in the context of Higgs searches~\cite{lepera}. The test statistic is
given by
\begin{equation}
  \label{eq:loglike2}
  {\cal{Q}}
  = -2\log \frac{L( \text{data}\, |\, {\text{Unhiggs~+~background}} ) } 
  { L( \text{data}\,| \, {\text{background}} ) }\,,
\end{equation}
where $L$ denotes the Poissonian likelihood, {\emph{e.g.}} 
\begin{equation}
L( \text{data}\, |\,  {\text{Unhiggs~+~background}}) =
 \frac{N^{n} e^{-N}}{n!}\,,
\end{equation}
where $N=(\sigma^{\rm{Unh}}+\sigma^{\rm{bkg}}){\cal{L}}$ is the number
of expected events at a given luminosity and $n$ is the number of
actually observed events in the Unhiggs model. The generalization to
binned histograms is straightforward.

The test statistic Eq.~\gl{eq:loglike2} is different from the profile
likelihood which is employed by ATLAS and CMS~\cite{Cowan:2010js} in
its asymptotic behavior and in the treatment of uncertainties. Since
the shape and systematic uncertainties are not publicly available, we
neglect them throughout and take the distributions by ATLAS and CMS at
face value. 

\begin{table*}[!t]
\begin{tabular}{|| c | c | c | c | c | c | c ||}
 \hline
 \multirow{2}{*}{Signal hypothesis} &  \multicolumn{3}{|c|}{95\% CL
   expected}  & \multicolumn{3}{|c||}{95\% CL observed}  \\
  &  \, ATLAS \,& \, CL$_S$ \, & \, MC+CL$_S$ \, & \, ATLAS
  \, & \, CL$_S$ \, & \, MC+CL$_S$ \, \\
  \hline
  $m_h=190\gev$ & 0.81 & 0.76 & 0.74 & 0.52 & 0.58 & 0.58\\
  $m_h =360\gev$& 0.79 & 0.76 & 0.76  & 0.57  & 0.56 & 0.57 \\  
 \hline
  \multirow{2}{*}{Signal hypothesis}  &  \multicolumn{3}{|c|}{95\% CL expected}  & 
 \multicolumn{3}{|c||}{95\% CL observed}  \\
 &  CMS & CL$_S$ & MC+CL$_S$ & CMS  & CL$_S$
 & MC+CL$_S$ \\
 \hline
$m_h=200\gev$ & 0.50 & 0.49 & 0.50 & 0.62 & 0.52 & 0.60\\
$m_h=350\gev$ & 0.73 & 0.67 & 0.67  & 0.76 & 0.78 & 0.73 \\
\hline
\hline
\end{tabular}
\caption{\label{tab:comp} Comparison of expected and observed 95\% 
  confidence level bounds for the various Higgs mass hypotheses of
  Fig.~\ref{fig:exiref}. We quote the numbers extracted from the ATLAS
  and CMS publications~\cite{atlaszz,cmszz}. ``CL$_S$'' gives the result
  of the hypothesis test with the histograms and the data of
  Fig.~\gl{fig:exiref} as input. ``MC+CL$_S$'' refers to the signal
  histograms generated with our Monte Carlo tool chain.}
\end{table*}

It can be expected that the influence of marginalization over nuisance
parameters~\cite{junk} on the computed confidence level is not too
important for this clean and well-reconstructible final state. Indeed,
we compute 95\% upper confidence levels using the CLs
method~\cite{Read:2002hq} in Tab.~\ref{tab:comp}, which are in good
agreement with the results from ATLAS and CMS. This holds especially
for large Higgs masses which we want to study in detail for the
purpose of this work. Given this agreement with ATLAS and CMS, our
implementation potentially reproduces the findings by ATLAS and CMS at
the percent level, well inside the $1\sigma $ uncertainty bands for
heavy Higgs bosons $m_h\gtrsim 225~\gev$.  For $m_h\lesssim 225~\gev$
our CL$_S$ implementation shows deviations of ${\cal{O}}(15\%)$ for
the experiments' histograms as input. We interpret this deviation as
the influence of systematics which plays an important role when the
$Z$ bosons are soft and off-shell. Since the resulting distributions'
shape uncertainties are not publicly known we cannot reproduce a
quantitatively reliable agreement for $m_h\lesssim 200~\gev$ within
the limitations of our naive detector simulation (for details see
below). This region requires the full experimental analysis flow,
which is not available to us. Hence, we focus in the following on the
heavy Higgses $m_h\gtrsim 200~\gev$.

\begin{figure*}[!t]
  \vspace{1cm}
  \subfigure[][\label{fig:unhiggscms}]{\includegraphics[width=0.4\textwidth]{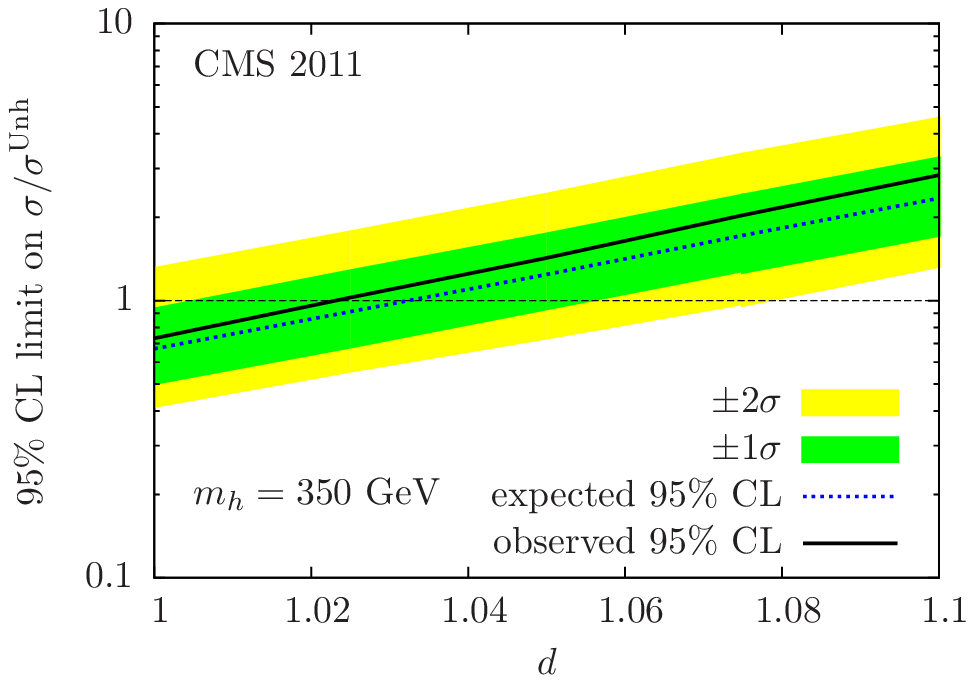}}
  \hspace{1cm}
  \subfigure[][\label{fig:unhiggsatlas}]{\includegraphics[width=0.4\textwidth]{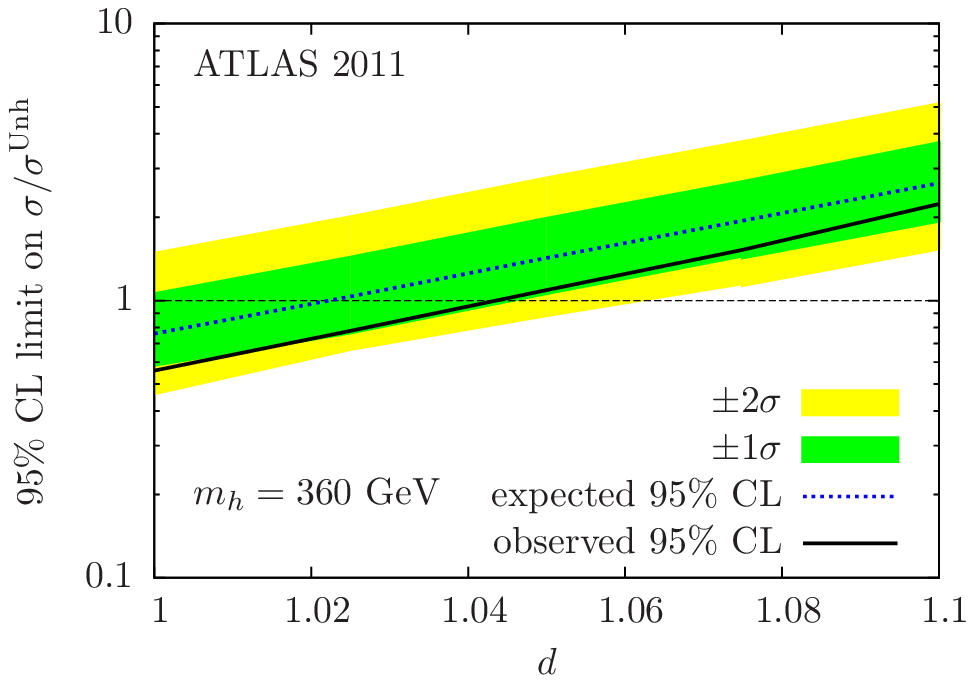}}\\
  \caption{\label{fig:unhiggscl} Observed (solid) and expected
    (dashed) 95\% confidence level exclusion for
    $\sigma/\sigma^{\rm{Unh}}$ for (a) CMS and (b) ATLAS. Note that
    for $d=1$ we recover the ``MC+CL$_S$'' values of
    Tab.~\ref{tab:comp}, {\emph{i.e.}}  the SM exclusion.}
\end{figure*}

\begin{figure}[!t]
  \includegraphics[width=0.4\textwidth]{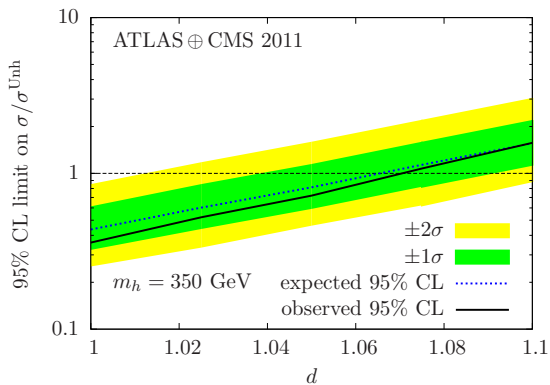}
  \caption{\label{fig:unhiggsclcomb} Combined observed (solid) and
    expected (dashed) exclusion by ATLAS and CMS for
    $\sigma/\sigma^{\rm{Unh}}$.}
\end{figure}

\subsubsection*{Event generation and MC Analysis}
For the event generation we use a modified version of MadGraph
v4~\cite{madgraph}, which implements the Unhiggs model as described in
Ref.~\cite{Englert:2012dq}. The parton level events are subsequently
showered with Pythia v6 \cite{pythia}. To account for detector
resolution effects we process the showered events with
PGS~\cite{pgs}. In order to reliably reproduce the SM Higgs signal
distributions that enter the hypothesis tests of ATLAS and CMS, we
have performed a dedicated tune of PGS, leading to good agreement of
the $m(4\ell)$ distribution at the percent level for heavy Higgs
masses $m_h\gtrsim 225~\gev$.

We adopt the cuts from the respective experimental
analysis~\cite{atlaszz,cmszz}.  Concretely this means selecting events
with two same-flavor and opposite-sign lepton pairs $\ell^{+}\ell^{-}$
with the following features required by each experiment:

\bigskip

\noindent \textbf{ATLAS} --- The leptons are required to have
transverse momenta of at least $p_{T}^{\ell}>7~\gev$, where two of
them need to pass from a further constraint of $p_{T}^{\ell}>20$
GeV. They are reconstructed within the geometrical coverage of
$|\eta_{e}|<2.47$ and $|\eta_{\mu}|<2.7$ with a separation $\Delta
R>0.1$. The first reconstructed $Z$ boson, $m_{Z_{1}}$, is required to
be on-shell by taking its invariant mass closest to the $Z$ boson mass
within the range $|m_{Z}-m_{Z_{1}}|<15~\gev$.  The invariant mass of
the remaining lepton pair, denoted by $m_{Z_{2}}$, is required to be
$m_{Z_{2}}^{\text{min}}<m_{Z_{2}}<115~\gev$. Where the threshold mass,
$m_{Z_{2}}^{\text{min}}$, depends on the reconstructed four-lepton
mass as denoted by the Tab.~\ref{tab:Atlas-threshold-m34}.\\

\bigskip 

\noindent \textbf{CMS} --- The leptons are required to have transverse
momenta of $p_{T}^{e}>7~\gev$ and $p_{T}^{\mu}>5~\gev$ being in the
pseudorapidity range of $|\eta_{e}|<2.5$ and $|\eta_{\mu}|<2.4$,
respectively. One of $Z$ bosons is reconstructed by the pair
$e^{+}e^{-}$ or $\mu^{+}\mu^{-}$ within the mass range $50~\gev
<m_{Z_{1}}<120~\gev$ and with the transverse momenta for the lepton
pair $p_{T,2e}>20~\gev$ or $p_{T,2\mu}>10~\gev$. The other $Z$ boson
is selected by the remaining same flavor combination
$\ell^{+}\ell^{-}$ and is denoted by $Z_{2}$. It is required that
$12~\gev<m_{Z_{2}}<120~\gev$ with the additional constrain
$m_{4\ell}>100~\gev$. If more than one combination satisfies these
criteria for $Z_{2}$ the one with the highest $p_{T}$
leptons is chosen.\\

\medskip 

We compare the resulting confidence levels in Tab.~\ref{fig:exiref},
where ``MC+CL$_S$'' denotes the hypothesis test with the background
hypothesis extracted from the experiments and with the signal
distributions generated by the described tool chain. In total we find
very good agreement for high masses, so that our analysis set-up is
sufficiently validated to confront the Unhiggs hypothesis with data

\begin{table}[t!]
  \begin{tabular}{||c|c|c|c|c|c|c|c|c|c||}
    \hline 
    $m_{4\ell}~\left[\gev\right]$  & $\leq$120 & 130 & 140 & 150 
    & 160 & 165 & 180 & 190 & $\geq$200\tabularnewline
    \hline 
    $m_{Z_{2}}^{\text{min}}~\left[\gev\right] $ & 15 & 20 & 25 & 30 
    & 30 & 35 & 40 & 50 & 60\tabularnewline
    \hline 
    \hline
  \end{tabular}
  \caption{\label{tab:Atlas-threshold-m34}  Threshold
    masses for $m_{Z_{2}}$  with $m_{4\ell}$ used in the ATLAS analysis.}
\end{table}

\section{Bounds on Unhiggs production from $ZZ\to 4\ell$}
\label{sec:bounds}
We show the expected and observed 95\% confidence level curves for the
high mass values quoted in Tab.~\ref{tab:comp} in
Fig.~\ref{fig:unhiggscl}. Throughout, we plot exclusion limits in
terms of $\sigma/\sigma^{\rm{Unh}}$ as function of the Unhiggs scaling
dimension~$d$, keeping $\mu=600~\gev$
fixed. $\sigma/\sigma^{\rm{Unh}}$ dominantly depends on $d$ for this
choice, and we discuss the influence of $\mu$ on the exclusion limits
in detail later on. We also show the {\hbox{ATLAS\,$\oplus$\,CMS}}
combination for $m_h=350~\gev$ in Fig.~\ref{fig:unhiggsclcomb}.

\begin{figure}[!b]
  \includegraphics[width=0.4\textwidth]{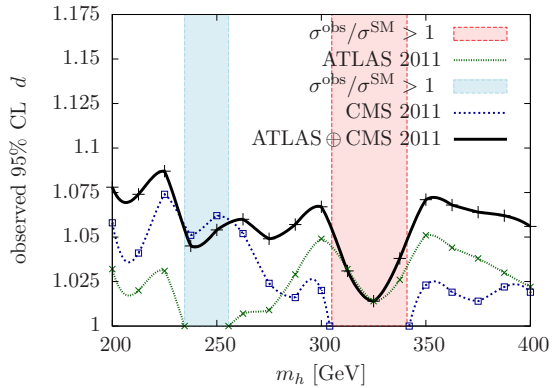}
  \caption{\label{fig:unhiggsdmh} Observed exclusion of the Unhiggs
    model ($\mu=600~\gev$) for ATLAS (green, dashed), CMS (blue,
    dotted) and the mark data excesses $\sigma/\sigma^{\rm{SM}}>1$, so
    that bounds on Unhiggs production cannot be imposed by ATLAS
    (blue) and CMS (red) individually.}
\end{figure}

\begin{figure}[t]
\includegraphics[width=0.38\textwidth]{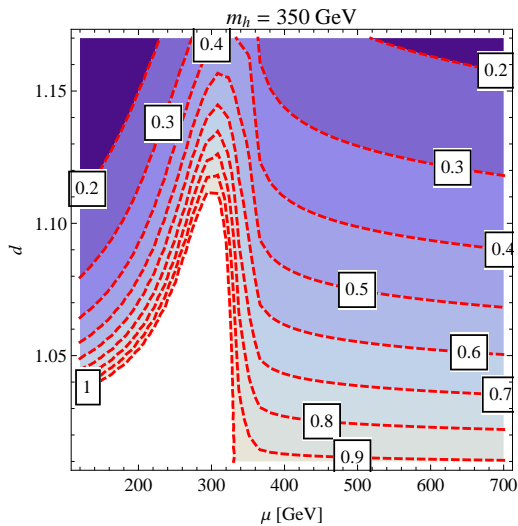}
\caption{\label{fig:mud} Representative
  $\sigma^{\rm{Unh}}/\sigma^{\rm{SM}}$ as a function of $d\geq 1.01$
  and $\mu$ for $m_h=350~\gev$. The blank area correponds to
  $\sigma^{\rm{Unh}}/\sigma^{\rm{SM}}>1$ and is already excluded.  For
  $d\leq 1.01$, $\sigma^{\rm{Unh}}/\sigma^{\rm{SM}}$ quickly
  approaches 1 from above. The observed exclusion by the combination
  coincides with $\sigma^{\rm{Unh}}/\sigma^{\rm{SM}}=0.5$.}
\end{figure}

In Fig.~\ref{fig:mud} we show the dependence of
$\sigma^{\rm{Unh}}/\sigma^{\rm{SM}}$ on $\mu$ for a representative
value of the Higgs mass. From this figure we see that a variation of
$\mu$ in the signal hypothesis leaves our findings for $d$ largely
unmodified unless we face a situation where the conformal symmetry
breaking scale $\mu$ is close to the Unhiggs pole mass. If we consider
the situation $\mu<m_h$ there is an additional contribution to the
width \cite{unhiggs}, which eventually can yield
$\sigma^{\rm{Unh}}/\sigma^{\rm{SM}}>1$. Consequently this region is
already now excluded at the 95\% confidence level by the combination,
and the observed limits on $d$ are slightly larger than for $\mu>m_h$,
with a stronger dependence on $\mu$.

The shape of the exclusion contours of Figs.~\ref{fig:unhiggscl} and
\ref{fig:unhiggsclcomb} is representative for the entire considered
mass range $200~\gev\leq m_h \leq 400~\gev$, while the quantitative
details of the observed exclusion on $d$ is mostly driven by actually
observed data around the $m_h$ signal hypothesis.  Local deficits and
excesses in the data do typically not change this situation,
{\emph{i.e.}} when we see a deficit in data this typically carries
over to a larger-than-expected $d$ for fixed $m_h,\mu$. The
combination of both searches in regions where the data is consistent
with the background allows to impose even stronger bounds on Unhiggs
production, as can be seen from the comparison of
Figs.~\ref{fig:unhiggscl} with \ref{fig:unhiggsclcomb}.

From Fig.~\ref{fig:exiref} (and Refs.~\cite{atlaszz,cmszz}), it is
clear that for some Higgs masses we observe excesses
$\sigma/\sigma^{\rm{SM}}>1$ by the individual experiments. Since the
Unhiggs model generically predicts smaller cross sections in $pp\to
ZZ+X$ for $\mu>m_h$ than encountered in the SM, we are therefore not
able to put bounds on the Unhiggs model in these particular regions
for the individual analyses.  CMS and ATLAS, however, observe these
excesses in different Higgs mass regions. This allows us to constrain
the full considered Higgs mass range in the Unhiggs model from the
combination of the ATLAS and CMS searches, yielding
$\sigma/\sigma^{\rm{SM}} < 1$ over the entire mass range $200~\gev\leq
m_h\leq 400~\gev$. This is shown in Fig.~\ref{fig:unhiggsdmh}, where
we scan the observed 95\% CL on $d$ over the mass range $200~\gev\leq
m_h \leq 400~\gev$ (again for $\mu=600~\gev$). The shaded areas
represent Higgs mass regions where ATLAS and CMS cannot constrain the
Unhiggs model individually. The combination of the two experiments
amounts to a combined bound of $d\sim 1.06$. In the region where no
bound can be imposed by the ATLAS experiment, the observed exclusion
is $\sigma/\sigma^{\rm{SM}}\simeq 1.5$, which results from a rather
large $\simeq 1.8\sigma$ upward fluctuation. This excess weakens the
observed CMS exclusion in the combination, which holds also vice-versa
for $m_h\simeq 325~\gev$ in a more pronounced way.

In total, the resulting SM bounds translate into only mild bounds on
Unhiggs production, {\it i.e.} $d\gtrsim 1.1$. Expecting
$d={\cal{O}}(1)$ in the Unhiggs scenario, these constraints are not
strong enough to rule out the existence of the Unhiggs scenario. This,
however, should be possible with the future increase of luminosity and
\hbox{center of mass energy}.

\vspace{0.5cm}
\section{Summary and Conclusions}
\label{sec:sum}
The LHC has tested Standard Model Higgs production at $7$ TeV center
of mass energy with a luminosity of about 5 fb$^{-1}$.  Significant
bounds on the SM Higgs could be established in 2011 by both ATLAS and
CMS. The combination of both data sets allows to impose constraints
$\sigma/\sigma^{\rm{SM}}<1$ over the range $200~\gev \lesssim m_h
\lesssim 400~\gev$ in the $pp\to ZZ+X\to 4\ell+X$ channel. We show
that we can reproduce the experiments sensitivity to very good
approximation, thus allowing us to understand the observed
underproduction if a Higgs is realized in this particular mass range
in terms of Unhiggs symmetry breaking. We find that the data only
mildly constrains the Unhiggs scenario $d\gtrsim 1.1$, with a very
flat dependence of these results on the high scale conformal symmetry
breaking scale as long as $\mu>m_h$.

\acknowledgements 
We cordially thank the organizers of the Heidelberg
New Physics Forum for a stimulating environment.
C.E. and M.S. thank the University of California Davis for the kind
hospitality during the time when parts of this work was
done. C.E. also would like to thank Markus Schumacher for helpful
discussions. C.E.  acknowledges funding by the Durham Junior Research
Fellowship scheme. D.G. acknowledges support by the International Max
Planck Research School for Precision Tests. J.T. was supported by the
Department of Energy under grant DE-FG02-91ER406746.

\end{document}